\documentclass[twoside,12pt]{article}
\usepackage{epsfig}

\newcommand{\be}{\begin{equation}}
\newcommand{\ee}{\end{equation}}
\newcommand{\bea}{\begin{eqnarray}}
\newcommand{\eea}{\end{eqnarray}}

\begin{document}

\title{ \vspace{1cm} Production of hypernuclei with hadronic and 
electromagnetic probes}
\author{R.\ Shyam$^{1,2}$ 
\\
$^1$Saha Institute of Nuclear Physics, Kolkata, India \\
$^2$Institute f\"ur Theoretische Physik, Universit\"at Giessen,   
Germany}
\maketitle
\begin{abstract} 
We present an overview of a fully covariant formulation of describing the 
hypernuclear production with hadronic and electromagnetic probes. This 
theory is based on an effective Lagrangian picture and it focuses on  
production amplitudes that are described via creation, propagation and 
decay into relevant channel of $N^*$(1650), $N^*$(1710) and $N^*$(1720)
intermediate baryonic resonance states in the initial collision of the
projectile with one of the target nucleons. The bound state nucleon and 
hyperon wave functions are obtained by solving the Dirac equation with 
appropriate scalar and vector potentials. Specific examples are discussed 
for reactions which are of interest to current and future experiments on 
the hypernuclear production. 
\end{abstract}
\section{Introduction}
Hypernuclei represent the first kind of flavored nuclei (with new quantum
numbers) in the direction of other exotic nuclear systems (e.g., charmed
nuclei). They introduce a new dimension to the traditional world of 
atomic nuclei. With a new degrees of freedom (the strangeness), they provide
a better opportunity to investigate the structure of atomic 
nuclei~\cite{gib95}. For example, since the $\Lambda$ hyperon does not 
suffer from the restrictions 
of the Pauli's exclusion principle, it can occupy all the states which are 
already filled up by the nucleons right upto the center of the nucleus. This 
makes it an unique tool to investigate the structure of the deeply bound 
nuclear states (see, e.g.,~\cite{has06,ban90}). The data on the
hypernuclear spectroscopy have been used extensively to extract information 
about the hyperon-nucleon ($YN$) interaction within a variety of theoretical 
approaches~\cite{hiy00,kei00}.

$\Lambda$ hypernuclei can be produced by beams of mesons, protons and also
heavy ions. Very recently, it has become possible to produce them also with
the electromagnetic probes like photons and electrons. Although, the stopped 
as well as in-flight $(K^-,\pi^-)$~\cite{chr89,ban90} and 
$(\pi^+,K^+)$~\cite{chr89,ban90,has06} reactions have been the most 
extensively used, the feasibility of producing hypernuclei via the $(p,K^+)$, 
$(\gamma,K^+)$ and $(e,e^\prime K^+)$ reactions has also been 
demonstrated in the recent years~\cite{kin98,shy06,yam95,iod07,yua06}. 

Several features of various hypernuclear production reactions can be 
understood by looking at the corresponding momentum transfers to the 
recoiling nucleus, since it controls to a great extent the population 
of the hypernuclear states. In Fig.~1, the momentum transferred to 
the recoiled nucleus is shown as a function of beam energy at two 
angles of the outgoing kaon for a number of reactions. We see that 
the $(K^-,\pi^-)$ reaction allows only a small momentum transfer to 
the nucleus (at forward angles), thus there is a large probability of 
populating $\Lambda$-substitutional states in the residual hypernucleus
($\Lambda$ occupies the same angular momentum state as that of the 
replaced neutron). On the other hand, in $(\pi^+,K^+)$ and 
$(\gamma,K^+)$ reactions the momentum transfers are larger than the 
nuclear Fermi momentum. Therefore, these reactions can populate states 
with configurations of a nucleon hole and a $\Lambda$ hyperon in a series 
of orbits covering all bound states. The momentum transfers
involved in the $(p,K^+)$ reaction are still larger by a factor of about 3.
Thus, the states of the hypernuclei excited in the $(p,K^+)$ reaction may have
a different type of configuration as compared to those excited in the
$(\pi^+,K^+)$ reaction. Nevertheless, it should be mentioned that 
usually larger momentum transfers are associated with smaller hypernuclear
production cross sections. Each reaction has its own advantage and plays its
own role in a complete understanding of the hypernuclear spectroscopy.
\begin{figure}[tb]
\begin{center}
\begin{minipage}[t]{8 cm}
\epsfig{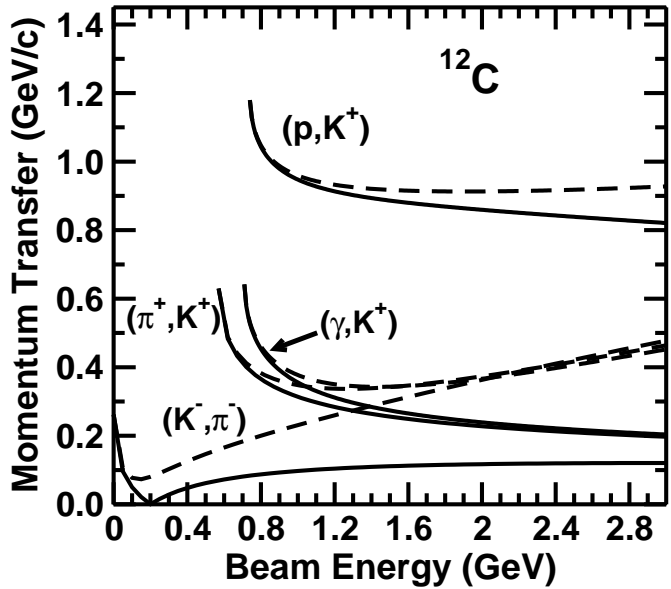}
\end{minipage}
\begin{minipage}[t]{16.5 cm}
\caption{The momentum transfer involved in various hypernuclear production
reactions as function of beam energy for the outgoing kaon angles of 
$0^\circ$ (full lines) and $10^\circ$ (dashed lines) for the $^{12}$C 
target.}
\end{minipage}
\end{center}
\end{figure}

With the recent successful completion of experiments at Jlab which produced
discrete hypernuclear states with electrons on $^7$Li and $^{12}$C targets
for the first time~\cite{iod07,yua06,miy03} and with more such experiments 
planned at Jlab and also at the accelerators MAMI-C and ELSA~\cite{poc05},
the exploration of the hypernuclei with electromagnetic probes has become of
great current interest. In contrast to the hadronic reactions $[(K^-,\pi^-)$ 
and $(\pi^+,K^+)]$ which take place mostly at the nuclear surface due to 
strong absorption of both $K^-$ and $\pi^\pm$, the $(\gamma,K^+)$ and 
$(e,e^\prime K^+)$ reactions occur deep in the nuclear interior since 
$K^+$-nucleus interaction is weaker. Thus, this reaction is an ideal tool 
for studying the deeply bound hypernuclear states if the corresponding 
production mechanism is reasonably well understood. While hadronic 
reactions excite predominantly the natural parity hypernuclear states, 
both unnatural and natural parity states are excited with comparable 
strength in the electromagnetic reactions~\cite{ben89,mot94,lee95,lee01}. 
This is due to the fact that sizable spin-flip amplitudes are present 
in the elementary photo-kaon production reaction, $p(\gamma,K^+)\Lambda$, 
since the photon has spin 1. This feature persists in the hypernuclear 
photo- and eletro-production. Furthermore, since in these reactions a 
proton in the target nucleus is converted into a hyperon, it leads to the
production of neutron rich hypernuclei (see, e.g., Ref.~\cite{bres05}) 
which may carry exotic features such as a halo structure.  It can produce 
many mirror hypernuclear systems which would enable the study
of the charge symmetry breaking with strangeness degrees of freedom.

\section{Covariant hypernuclear production amplitudes}

Since we are still far way from calculations of the intermediate energy 
scattering and reactions directly from the lattice QCD, the effective field
theoretical description in terms of the baryonic and mesonic degrees
of freedom, is usually employed to describe these processes. These approaches
introduce the baryonic resonance states explicitly in their framework and
QCD is assumed to provide justification for the parameters or the 
cut-off functions used in calculations. 

We use the diagrams shown in Fig.~2 for the calculations of the proton, meson
and photon induced hypernuclear production reactions. In all the three cases,
the initial state interaction of the projectile with a bound nucleon of 
the target leads to excitations of $N^*$(1650)[$\frac{1}{2}^-$], 
$N^*$(1710)[$\frac{1}{2}^+$], and $N^*$(1720)[$\frac{3}{2}^+$] baryonic 
resonance intermediate states which decay into kaon and the $\Lambda$ 
hyperon which gets captured into one of the nuclear orbits. These resonances 
have appreciable branching rations for the decay into the $K^+\Lambda$ 
channel and are known to contribute predominantly to the corresponding 
elementary reactions involved in various processes~\cite{shy99,shk05}.   
\begin{figure}[tb]
\begin{center}
\begin{minipage}[t]{8 cm}
\epsfig{file=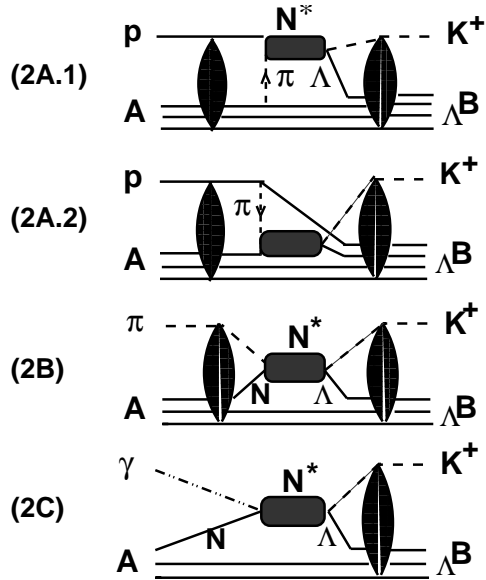,scale=0.5}
\end{minipage}
\begin{minipage}[t]{16.5 cm}
\caption{Types of the Feynman diagrams included in calculations of 
various reactions. In case of the $(p,K^+)$ reaction two types of 
diagrams, the target emission (Fig. 2A.1) and projectile emission 
(Fig. 2A.2) contribute to the total amplitude. The elliptic shaded areas 
represent the optical model interactions in the incoming or outgoing 
channels}
\end{minipage}
\end{center}
\end{figure}

To evaluate the amplitudes corresponding to the diagrams shown in Fig.~2, 
we require effective Lagrangians for the hadronic and electromagnetic
couplings of the resonances. These are described in Refs.~\cite{shy06,shy07}.
The coupling constants at various interaction vertices, propagators for the
intermediate resonance states and the form factors for the 
resonance-nucleon-meson vertices are all described in 
Refs.~\cite{shy06,shy07,shya06}. In the case of the $(p,K^+)$ reaction the
initial interaction between the incoming proton and a bound nucleon of the 
target is modeled by means of $\pi$, $\rho$ and $\omega$ exchange mechanisms 
as discussed in Ref.~\cite{shy06}. Terms corresponding to the interference 
between various amplitudes are retained in the production amplitudes. Since 
calculations within this theory are carried out in the momentum
space all along, they includes all the nonlocalities in the production 
amplitude that arises from the resonance propagators. 
\begin{figure}[tb]
\begin{center}
\begin{minipage}[t]{8 cm}
\epsfig{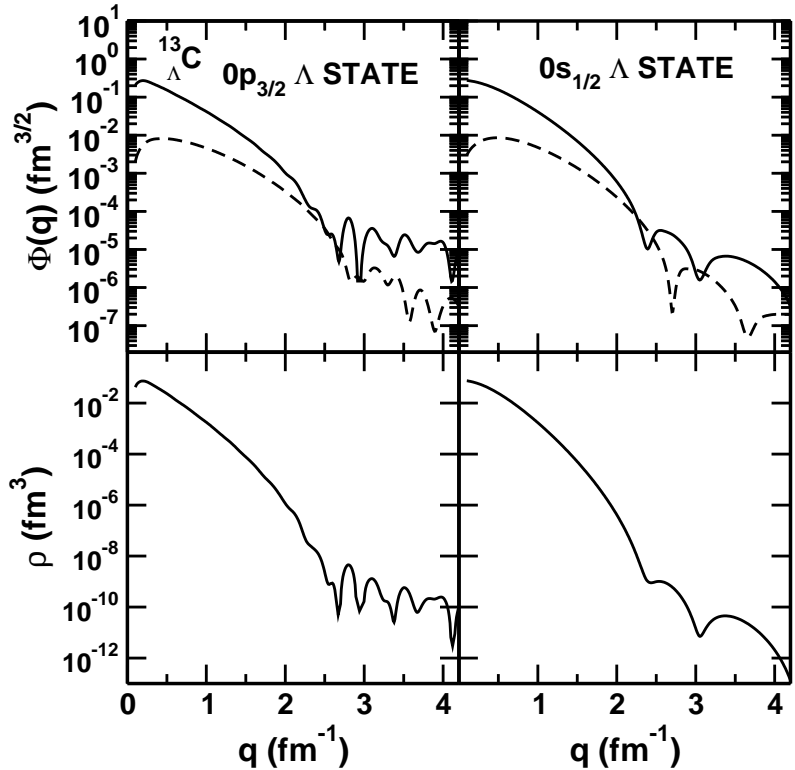}
\end{minipage}
\begin{minipage}[t]{16.5 cm}
\caption{
(Upper panel) Momentum space spinors [$\Phi(q)$] for the $0p_{3/2}$
and $0s_{1/2}$ $\Lambda$ orbits in $^{13}\!\!\!_\Lambda$C hypernucleus.
$|F(q)|$ and $|G(q)|$ are the upper and lower components of the spinor,
respectively. (Lower panel) Momentum distribution [$\rho(q) = 
|F(q)|^2 + |G(q)|^2$] for the same hyperon state calculated with these 
wave functions.
}
\end{minipage}
\end{center}
\end{figure}

A crucial input required in the calculations of the $(p,K^+)$ hypernuclear 
production reaction is the pion self-energy [$\Pi(q)$] which takes into 
account the medium effects on the intermediate pion propagation. Since the 
energy and momentum carried by such a pion can be quite large (particularly 
at higher proton incident energies), the calculation of $\Pi(q)$ within a 
relativistic approach is mandatory. In our calculations of $\Pi(q)$,
we take into account the contributions from the particle-hole ($ph$) and 
delta-hole ($\Delta h$) excitations produced by the propagating 
pions~\cite{shy04}. The self-energy has been renormalized by including 
the short-range repulsion effects through the constant Landau-Migdal 
parameter $g^\prime$ which is taken to be the same for $ph-ph$ and 
$\Delta h-ph$ and $\Delta h-\Delta h$ correlations which is a common 
choice. $g^\prime$ acting in the spin-isospin channels, is supposed 
to mock up the complicated density dependent effective 
interaction between particles and holes in the nuclear medium. Most estimates
give a value of $g^\prime$ between 0.5-0.7.

\section{Results and Discussions}

The spinors for the final bound hypernuclear state and for intermediate 
nucleonic states are required to perform numerical calculations of various 
amplitudes. We assume these states to have pure-single particle or 
single-hole configurations. The spinors in the momentum space are
obtained by Fourier transformation of the corresponding coordinate space
spinors which are the solutions of the Dirac equation with potential fields
consisting of an attractive scalar part ($V_s$) and a repulsive vector part
($V_v$) having a Woods-Saxon form. The parameters of these potentials  
are given in Refs.~\cite{shy06,shy07}.

In Fig.~3, we show the lower and upper components of the Dirac 
spinors in momentum space for $0p_{3/2}$ and $0s_{1/2}$ hyperons in
$^{13}\!\!\!_\Lambda$C. In each case, we note that only
for momenta $<$ 1.5 fm$^{-1}$, is the lower component of the spinor 
substantially smaller than the upper component. In the region of momentum
transfer pertinent to exclusive kaon production, the lower components of 
the spinors are not negligible as compared to the upper component. This 
clearly demonstrates that a fully relativistic approach is essential for
an accurate description of this reaction. The spinors calculated in this 
way provide a good description of the experimental nucleon momentum 
distributions for various nucleon orbits as is shown in Ref.~\cite{shy95}.
In the lower panels of Fig.~3 we show momentum distribution, 
$\rho(q)\,[ = |F(q)|^2 + |G(q)|^2]$, of the corresponding $\Lambda$ hyperon.
In each case the momentum density of the hyperon shell, in the momentum 
region around 0.35 GeV/c, is at least 2-3 orders of magnitude larger than 
that around 1.0 GeV/c. Thus reactions involving lower momentum transfers
are expected to have larger cross sections. 

\begin{figure}[tb]
\begin{center}
\begin{minipage}[t]{8 cm}
\epsfig{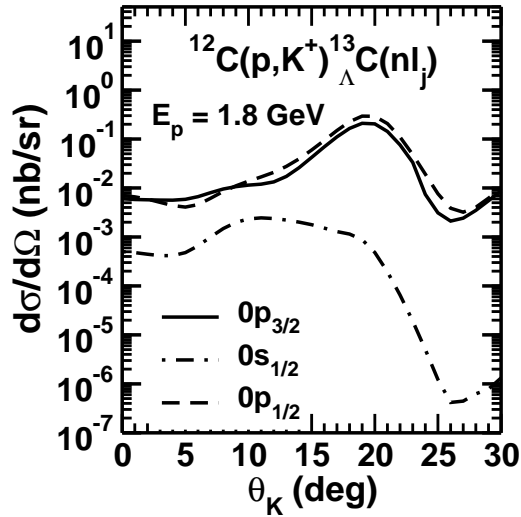}
\end{minipage}
\begin{minipage}[t]{16.5 cm}
\caption{
Differential cross section for the
$^{12}C(p,K^+)^{13}\!\!\!_\Lambda C$
reaction for the incident proton energy of 1.8 GeV for various
bound states of final hypernucleus as indicated in the figure.
}
\end{minipage}
\end{center}
\end{figure}

In Fig.~4,  we show the kaon angular distributions for various final
hypernuclear states excited in the reaction 
$^{12}$C$(p,K^+)$$^{13}\!\!\!_\Lambda$C. We have used the plane waves to
describe the scattering wave functions in the initial and final channels. 
The incident proton energy  is taken to be 1.8 GeV where the angle integrated
cross sections for this reaction has a maximum. We have used  $g^\prime$ = 
0.5 in these calculations. The results are the coherent sum of all the 
amplitudes corresponding to the various meson exchange processes and 
intermediate resonant states. Clearly, the cross sections are quite selective 
about the excited hypernuclear state, being maximum for the state of largest 
orbital angular momentum. This is a direct reflection of the large momentum 
transfer involved in this reaction. 

We note that the angular distributions have a maximum at angles larger 
than 0$^\circ$. This can be understood by noticing that the bound state 
spinors of $^{13}\!\!\!_\Lambda$C, have several maxima in the upper and 
lower components in the region of large momentum transfers. Therefore, in 
the kaon angular distribution the first maximum may shift to larger angles 
reflecting the fact that the bound state wave functions show diffractive 
structure at higher momentum transfers. Furthermore, there is a 
difference of more than an order of magnitude between the cross sections for
the $p$-shell and the $s$-shell excitations. This is a direct reflection of 
the fact that the binding energies ($\varepsilon $) of the $p$-shell states 
are in the range of only 0.7-0.9 MeV as compared to 11.70 MeV of the $s$-shell.
The bound state spinors behave at larger momentum transfers, approximately 
as $\frac{1}{k^2}$ with $k \propto \varepsilon$. Thus, with increasing binding
energies the magnitude of $\Phi$ decreases, leading to the
corresponding decrease in the cross sections. Therefore, the $p$-shell 
transitions are enhanced as compared to the $s$-shell one.
 
In Fig.~5, we show the dependence of our calculated cross sections on
pion self-energy. It is interesting to note that this has a rather large
effect. We also note a surprisingly large effect on the short range 
correlation (expressed schematically by the Landau- Migdal parameter 
$g^\prime$). Similar results have also been reported in case of the 
$(p,\pi)$ reactions~\cite{shy95}.  
\begin{figure}[tb]
\begin{center}
\begin{minipage}[t]{8 cm}
\epsfig{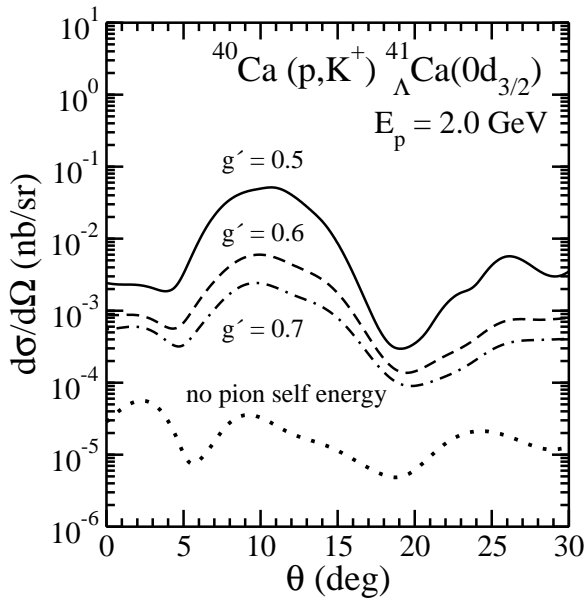}
\end{minipage}
\begin{minipage}[t]{16.5 cm}
\caption{
Differential cross section for the
$^{40}$Ca$(p,K^+)$$^{41}\!\!\!_\Lambda$Ca($0d_{3/2}$)
reaction for the incident proton energy of 2.0 GeV.
The dotted line shows the results obtained without including the pion
self-energy in the denominator of the pion propagator while
full, dashed and dashed-dotted lines represent the same calculated
with pion self-energy renormalized with Landau-Migdal parameter of
0.5, 0.6 and 0.7, respectively.
}
\end{minipage}
\end{center}
\end{figure}
  
In Fig.~6, we present the total cross sections for populating 
$(0p_{3/2}^{-1},0p_{3/2}^\Lambda) 3^+, 2^+,0^+ $ states in the
$^{16}$O$(\gamma,K^+)$$^{16}\!\!\!_\Lambda$N reaction as a function of
photon energy. The contributions of all the resonances are coherently
summed in the depicted cross sections. The distortion effects are ignored 
in the $K^+$ channel; these effects are weak for $(\gamma,K^+)$ reaction
on lighter targets~\cite{ben89,ros88}.  A noteworthy aspect of this figure
is that cross sections peak at photon energies around 900 MeV, which is 
about 200 MeV above the production threshold for this reaction. Interestingly,
the total cross section of the elementary $p(\gamma,K^+)\Lambda$ reaction 
also peaks about the same energy above the corresponding production threshold
(~910 MeV). Cross sections near the peak position are about 20 nb which 
should be measurable at MAMI-C and ELSA accelerator facilities. 

We see in Fig.~6 that the highest $J$ state is most strongly excited. 
From kinematical considerations it is easy to note that for large momentum 
transfer the orbital angular momentum transfer will also be large. 
Furthermore, we note that the cross section of the unnatural parity state 
$3^+$ is larger than that of the natural parity state $2^+$  
by about a factor of 2.5 and by more than an order of magnitude than that 
of the $0^+$ state. The unnatural parity states are excited through the 
spin flip process. Thus, kaon photoproduction on nuclei is an ideal tool 
for probing the structure of the unnatural parity hypernuclear states. 
The addition of unnatural parity states to the spectrum of hypernuclei is 
expected to constrain the spin dependent part of the effective 
$\Lambda-N$ interaction more tightly.
\begin{figure}[tb]
\begin{center}
\begin{minipage}[t]{8 cm}
\epsfig{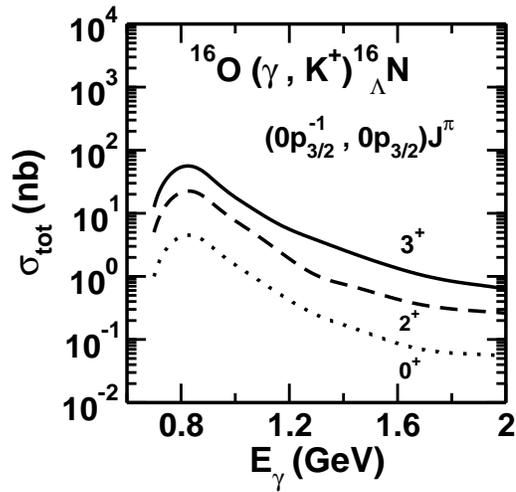}
\end{minipage}
\begin{minipage}[t]{16.5 cm}
\caption{
Total cross section for the $^{16}O(\gamma,K^+)^{16}\!\!\!_\Lambda N$
reaction as a function of photon energy. Results are shown for various
states corresponding to the $(0p_{3/2}^{-1},0p_{3/2}^\Lambda)$
configuration.
}
\end{minipage}
\end{center}
\end{figure}
Within our formalism, the measured differential cross sections for the
$(\gamma,K^+)$ reaction can be used to probe directly the bound $\Lambda$ wave 
functions provided the distortion effects in the kaon channels are weak 
and the nucleon bound state wave function is well controlled from the 
$(e,e^\prime p)$ data. Such information may prove to be very useful
in certain cases where adding a $\Lambda$ particle to the nucleus can
lead to significant rearrangements in the structure of the nucleus.
Such a possibility is discussed in Ref.~\cite{hiy96}. 

We find that both $(\gamma,K^+)$ and $(p,K^+)$ reactions are dominated by 
the excitation of the positive parity spin-$\frac{1}{2}$ $N^*$(1710) 
intermediate resonance state. The contributions of spin-$\frac{3}{2}$ 
$N^*$(1720) state is almost negligible in both the cases 
(see also Refs.~\cite{shy04,shy07}).

\section{Summary and Conclusions}
In summary, it is clear that the hypernuclear spectroscopy is indispensable
for the quantitative understanding of the $\Lambda$ hypernuclear structure
and the $\Lambda N$ interaction. Driven by the new proton and electron
accelerator facilities, the field of hypernuclear production with protons, 
photons and electrons is expected to experience a grand revival. Already
some data on the electron induced reactions are available from JLab and
data on photon induced reactions are expected to be available in near
future from MAMI-C and ELSA facilities. At the same times, our understanding 
of the hypernuclear production mechanism has improved significantly
over the last decade.

A fully relativistic approach is essential for an accurate description of
the hypernuclear production cross sections. It is feasible to calculate 
the reactions induced by  hadronic and electromagnetic probes within a 
single fully covariant effective Lagrangian picture. Since the relevant 
elementary production cross sections
are also described within the similar picture, most of the input parameters
needed for the calculations of the hypernuclear production are fixed 
independently.

The calculated $(p,K^+)$ cross sections are maximum for the hypernuclear 
state with the least binding energy and largest orbital angular momentum. 
The angular distributions for the favored transitions peak at angles larger
than $0^\circ$ which in contrast to the results of most of the previous
nonrelativistic calculations for this reaction. This reflects
directly the nature of the Dirac spinors for the bound states which involve
several maxima in the region of large momentum transfers. The 
nuclear medium corrections to the intermediate pion propagator
introduce large effects on the kaon differential cross sections. There is also
the sensitivity of the cross sections to the short-range correlation parameter
$g^\prime$ in the pion self-energy. Thus, $(p,K^+)$ reactions may provide an
interesting tool to investigate medium corrections on the pion propagation
in nuclei.

The $(\gamma,K^+)$ reaction selectively excites the high spin
unnatural parity states. Thus kaon photoproduction on nuclei is an ideal
tool for investigating the spin-flip transitions which are only weakly
excited in reactions induced by hadronic probes. Therefore, electromagnetic
hypernuclear production provides a fuller knowledge of hypernuclear
spectra which are used to investigate the details of the effective 
hyperon-nucleon interaction in the nuclear matter. A complete 
information about this spectrum will impose more severe constraints on the 
models of the $YN$ interaction, particularly on its poorly known spin 
dependent part. Therefore, measurements for this reaction at future electron 
accelerator facilities would have very exciting prospects.

The author wishes to thank Horst Lenske and Ulrich Mosel for their 
collaboration on this subject. This work has been supported by the 
Sonderforschungsbereich/Transregio 16, Bonn-Giessen-Bochum of the German 
Research Foundation. One of the authors (RS) acknowledges the support of 
Abdus Salam International Centre for Theoretical Physics in form of a senior 
associateship award.

\end{document}